\documentclass[12pt,a4paper]{article}
\usepackage{graphicx}
\usepackage{amsmath}
\usepackage{amssymb}
\usepackage{amsbsy}
\usepackage{makeidx}
\usepackage{setspace}
\usepackage{framed}

\makeindex
\date{}

\begin{document}
\setcounter{tocdepth}{5}
\setcounter{secnumdepth}{5}
\setcounter{page}{6}
\newpage
\pagenumbering{arabic}
\setcounter{page}{1}
\large
\baselineskip25pt
\title{Poincar\'{e} parameters and principal component analysis of Heart rate variability of subjects with health disorder}
\author{\\[0.2in]Sobhendu Kumar Ghatak${}^1$\\Subhra Aditya${}^2$\\ ${}^{1}$ Department of  Physics, Vidyamandir, Belur, 700123, India \\ ${}^{2}$ Department of Cardiology, R.G.Kar Hospital Kolkata,700004, India}

\maketitle
\newcommand{\be}{\begin{equation}}
\newcommand{\ee}{\end{equation}}
\newcommand{\ba}{\begin{eqnarray}}
\newcommand{\ea}{\end{eqnarray}}
\def\Eins{{\mathchoice {\rm 1\mskip-4mu l} {\rm 1\mskip-4mu l}
           {\rm 1\mskip-4.5mu l} {\rm 1\mskip-5mu l}}}

\maketitle
\begin{abstract}
Heart rate variability,important marker for modulation of autonomic nervous system is studied for diabetic,hypertensive and control group of subjects.Lagged Poincar\'{e} plot of heart rate(HR),method of principal component analysis and auto-correlation of HR fluctuation are used to analyze HR obtained from ECG signal recorded over short time duration.The parameters $(SD1)$,$(SD2)$ and their ratio $(SD12)$,characterizing the Poincar\'{e} plot reveal a significant reduction of their values for diabetic and hypertensive subjects compared to the corresponding results of control one.The slope and the curvature of the plot of these parameters with lagged number exhibit similar trend.In particular,the curvature of $(SD12)$ for the control group differs widely from that of other groups.The principal component analysis is used for analysing multi-dimensional data set resulting from the Poincar\'{e} plot for all subjects.The correlation matrix points out significant correlation between slope and curvature.The analysis demonstrates that three groups are well separated in the domain of two significant principal components.The auto-correlation of HR fluctuation exhibits highly correlated pattern for subjects with health disorder compared to that of control subject.
\end{abstract}
\vspace{0.2cm}
\noindent\textbf{Keywords}: Heart Rate Variability;Poincar\'{e} analysis;Principal Components;Diabetic;Hypertensive.\\
\textit{e-mail for correspondence} :skghatak@phy.iitkgp.ernet.in

\newpage
\section{Introduction}

The heart rate (HR) is obtained from the measurement of two consecutive R-R interval in ECG signals and is regulated predominantly by autonomous nervous system(ANS).The variability in HR,coined as heart rate variability (HRV),is a measure of the RR-interval fluctuations that originates from rhythmic cardiac activity. HRV is primarily controlled through non-additive activity of the sympathetic and parasympathetic branches of ANS.Sympathetic response increases the heart rate whereas parasympathetic does the reverse. Any disorder in health condition resulting in dysfunction of autonomous nervous system would alter heart rate and will be reflected in the change of HRV. The measurement of heart rate variability(HRV) being non-invasive,sensitive,faster and reproducible,is most widely utilized to assess cardiovascular autonomic dysfunction \cite {task} -\cite {Acharya}.A number of indices of HRV obtained from the detailed and sophisticated analysis seem to establish notable relationship between ANS and various diseases \cite {Eberhard}\cite{Sztajzel}. A decreased HRV values represent a well-known marker of potential cardiovascular risk for these patients where heart rate is reduced due to ANS dysfunction \cite {Levy}.A disfunction is often prevalent to patients with prolonged illness like hypertension \cite{Madsen},\cite{Langewitz} and diabetic \cite {Pagani}, \cite {Schroeder}, \cite {Javorka}.\\

The HRV is traditionally quantified from HR data using linear measures in the time and frequency domain.The data acquired for long time provides better measure in frequency domain. However, it is  inconvenient for subjects for long time ECG recording in a given position and also time consuming.The spectral analysis in frequency assumes stationarity of signal and sudden changes in HR results power spectrum which often difficult to interpret.The power spectrum of HR signal provides the power in high frequency (HF) band ($ 0.15$-$0.4 $ )Hz representing mostly of parasympathetic response and low frequency (LF) band ($ 0.04-0.15$ )Hz resulting from combined influence of sympathetic and that (albeit small) parasympathetic responses.The ratio of power LF/HF is often considered as a measure of sympathovagal balance \cite {task}.Simpler method to assess complex non-linear behavior in the study of physiological signals is Poincar\'{e} plot \cite{Eberhard}, \cite{Kamen},\cite{Piskorski} .
The Poincar\'{e} plot,often used in the study of non-linear dynamics, depicts graphically HR fluctuation and it is scattered plot where each RR interval is plotted against its next interval.
The plot uses unfiltered data and simpler to study dynamics of heart variability.The scattered plot of RR intervals forms a cluster and the visual inspection of the shapes of the cluster in the Poincar\'{e} plot is a helpful to find the quality of recorded ECG signals and identification of premature and ectopic beats \cite{Mourot},\cite{Sosnowski}.The studies based on short-term HRV with epochs as short as $300$ beat to beat interval (around $5$ minutes of ECG recording) in time domain suggest that the measures are better reproducible than frequency domain measures \cite {Sztajzel}.The measures are different in two domains but are not mutually independent due to strong correlation between the parameters.As a result any additional information useful for better discrimination of subjects with cardiovascular deregulation are limited \cite {Liao}.Therefore,the search for new parameters,which are able to provide additional information embedded in the HRV signals is important area of research.In the plot it is implicitly assumed that two successive R-R intervals are well correlated.This assumption lends itself to further penalization to lagged Poincar\'{e} plots by plotting $RR_{i+m}$ against $RR_{i}$ where $m$ represents the distance (in number of beats) between duplet beats.It is assumed that the heart is controlled by a nonlinear deterministic system and therefore nonlinear analysis of HRV is expected to provide better information \cite{Eberhard}, \cite{Bergfeldt},\cite{Voss}.The Lagged Poincar\'{e} method,an extension of conventional Poincar\'{e} one has been found to be a better quantitative tool  \cite {Karmakar} and is substantiated from some studies on Chronic Renal failure (CRF) subject populations \cite{Lerma},Congestive Heart Failure (CHF)\cite{Thakre}, diabetic subjects \cite{Contreras},\cite{Bhaskar} and rotatory audio stimulation \cite{Bhaskar Roy}.Different methodology for non-linear analysis of HRV are being evolved \cite{Voss}. The methods like 3-D return mapping \cite{Ruy}, wavelet analysis \cite{Acharya},detrended fluctuation analysis (DFA) \cite{Zhi}, \cite{Hu}, \cite{{Eduardo}} have been applied to extract different aspects of variability of heart rate.\\
In this work,the heart rate variability of diabetic, hypertensive and control subjects are studied in time domain from ECG signal of short duration.The variability are analyzed using measures obtained from  the lagged Poincar\'{e} plot,the Principal component analysis (PCA) and the correlation of HR fluctuation.The study is aimed to explore possibility of finding a specific outcome that differentiate three groups easily.

\section{Method}
Subjects were chosen from the out-patient department of SSKM/IPGMER Hospital,Kolkata as per inclusion and exclusion criteria of the study.Each of them was explained about the object of study,and written consent was taken.Permission of the Institutional ethical committee of the hospital was taken.The total number of subjects in this project was $118$ and was divided into three groups 1) control  (C), 2) diabetics (DM) and 3) hypertensive (HTN) with the population   $49$,$34$ and $35$ respectively.Most of the subjects came from low income social strata.The average age of HTN group was $55 \pm 19$ years whereas that of others were in range $47 \pm 18$ years.The ECG data were recorded in supine position for $6$ minutes duration with lead II configuration at the sampling rate of $500$Hz.The RR interval were extracted from the ECG data with the help of ORIGIN software.The peaks other than regular one were discarded in the analysis.\\
\subsection*{Poincar\'{e}Plot}


The Poincar\'{e} plot is a scatter plot of $RR_{i}$ vs. $RR_{i+1}$ where $RR_{i}$ is the time interval between two consecutive $R$ peaks and $RR_{i+1}$ is the time between the next two successive $R$ peaks \cite {task},\cite {Sztajzel},\cite{Bergfeldt},\cite{Kamen},\cite {Karmakar}.Apart from this conventional plot $(RR_{i+1}$ vs. $RR_{i})$, we also used the generalized Poincar\'{e} plot with different intervals,including the $m$-lagged Poincar\'{e} plot (the plot of $RR_{i+m}$ against $RR_{i}$).The plot is fitted with an ellipse with semi-major axis as bisector of plot axes.On the  plot,the semi-minor axis is the width and the semi-major part is the measure of the length of the ellipse. The ellipse-fitting technique provides three indices: width,length and their ratio which are respectively referred as $(SD1)$,$(SD2)$ and $SD12 = SD1/SD2$. The parameter $(SD1)$ is the measure of the standard deviation of instantaneous beat-to-beat interval variability and $(SD2)$ is that of the continuous long-term RR interval variability.The ratio $SD12 = SD1/SD2$ is the relative measure.The lagged-Poincar\'{e} plot provides these parameters for different lag $m$.All parameters are an increasing function of $m$.The growth of the parameters with $m$ can be quantified by the slope and the curvature of the respective curve and as their values are variable the largest values that occurred near $m=1$ are most relevant for characterization of growth with $m$.

\subsection*{Principal Components Analysis}
Principal component analysis (PCA) is a statistical procedure by which a large set of correlated variables can be transformed to a smaller number of independent new set of variable without throwing out essence of original data set \cite{Jolliffe}.The new set of variables are referred as the principal components (PCs).They are linear combination of original one with weights that form orthogonal basis vectors.These basis vectors are eigenvectors of covariant or correlation matrix of data set.Each component carries new information
about the data set,and is ordered so that the first
few components sufficiently provide most of the variability.The PCA is useful procedure for data reduction,visualization and feature extraction from multidimensional data.
In this study the PCA is applied to the parameters obtained from extended Poincar\'{e} plot \cite {Jolliffe},\cite {Mika}.The variables namely $SD1$,$SD2$,$SD12$ for lag $m=1$ and maximum value of the slope and the curvature of $SD1$,$SD2$,$SD12$  vs $m$ are considered assuming that these parameters specify important features of the Poincar\'{e} plot.As the parameters have different units it is imperative to normalize their values to put them so in same footing.The normalization was done so that the median of each parameter of data set was $0$ and the half-width of distribution was $0.5$.The dimensionality of data is therefore $9$.The data set is presented as matrix $X_{M,N}$ where there are $M$ number of subject and $N$ number of parameters of the data and is given as
\begin{equation}\label{eq.2}
  X_{M,N} =
 \begin{pmatrix}
  x_{1,1} & x_{1,2} & \cdots & x_{1,N}\\
  x_{2,1} & x_{2,2} & \cdots & x_{2,N}\\
  \vdots  & \vdots  & \ddots & \vdots \\
  x_{M,1} & x_{M,2} & \cdots & x_{M,N}
   \end{pmatrix}
\end{equation}
The elements $x_{i,j}$ represents the $j$th parameter value of $i$th subject.The covariance matrix $C$ of normalized data was then obtained as
\begin{equation}
  C = \frac{1}{M - 1}X^{T}X
  \label{matrix}
\end{equation}

The eigenvalues and the eigenvectors of matrix $C$ are obtained for each group using MATLAB programme.Out of $N$ eigenvalues it turns out that the first $n$ (according to their magnitudes) values can account most of the data and the corresponding $n$ eigenvectors are considered as basis vectors for calculation of principal components.The eigenvector is given by

\begin{equation}
 \Psi_{N,n} =
 \begin{pmatrix}
  \varphi_{1,1} & \varphi_{1,2} & \cdots & \varphi_{1,n}\\
  \varphi_{2,1} & \varphi_{2,2} & \cdots & \varphi_{2,n}\\
  \vdots  & \vdots  & \ddots & \vdots \\
  \varphi_{N,1} & \varphi_{N,2} & \cdots & \varphi_{N,n}
   \end{pmatrix}
 \label{eigvec}
\end{equation}
The principal components $PC$ can then be calculated from the  matrix equation
\begin{equation}\label{eq.4}
  PC_{n,M} = \Psi^{T}X^{T}
\end{equation}
where  $PC$ is the matrix of {nxM}.The eigenvector is the measure of the weight of data in determining the principal components and so $PC_{n,M}$ is $n$ principal component of $M$ subject.
The elements of covariant matrix represent the amplitude of correlation between the Poincar\'{e} parameters.
\subsection*{Correlation Function}
In order to estimate the strength of the correlation and stationarity of the heart rhythm the time-series of the increments of $RR_{i}$) was constructed and the correlation of deviation of successive beat interval was calculated.
The auto-correlation function $Corr(m)$ was obtained from the equation
\begin{equation}
Corr(m) = \frac{1}{\sigma} \sum_{i}^{N}\Delta RR_{i}\Delta RR_{i-m}
\label{cor}
\end{equation}
where $\sigma$ is the normalizing factor $Corr(0)$ and $\Delta RR_{i} = RR_{i+1} - RR_{i}$ is the deviation between two consecutive beat interval.

\subsection{Results and Discussions}
\subsubsection*{\textit{The Poincar\'{e} Parameters}}

A representative scattered Poincar\'{e} plot for control (C),hypertensive (HTN) and diabetic (DM) subjects with similar age are depicted in Fig.$1$ for all three groups for $m=1$ and $m=9$.Each point represents a $RR$ interval.The left side figures are plots of two successive intervals whereas the points on the right side figures represent the beat intervals are separated by $9$ successive interval.Note that with lag the points are scattered in a nearly random way.

\begin{figure}[h!]
\centering
\includegraphics[width=1.1\textwidth]{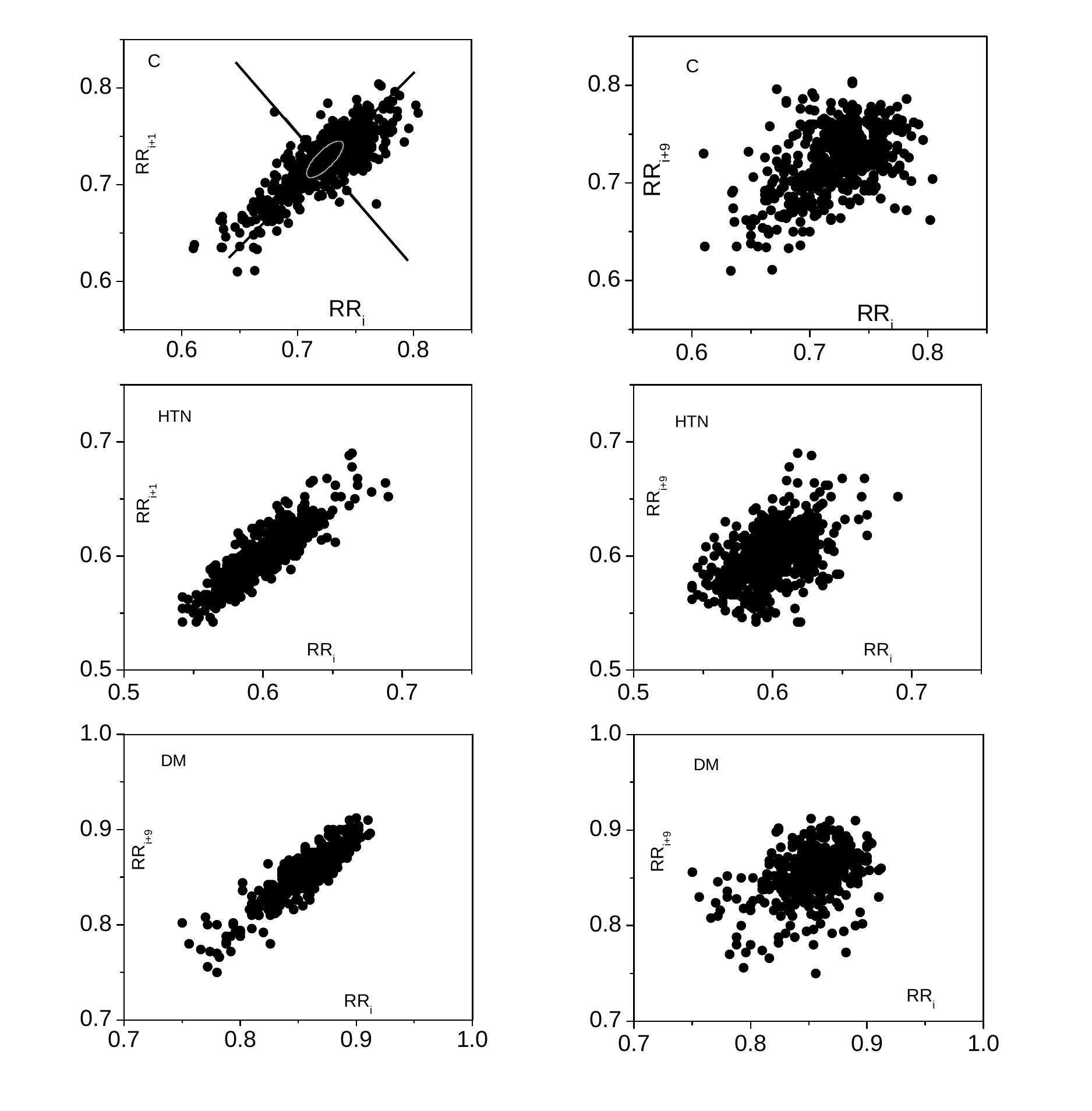}
\caption{The Poincar\'{e} plot for control (C),hypertensive (HTN) and diabetic (DM) subjects for $m=1$ and $m=9$.The length of semi-minor and semi-major axes of the representative ellipse in plot $RR_{i+1}$, $RR_{i}$ of control subject determines the parameters $(SD1)$ and $(SD2)$ respectively.}
\label{PoincareFinal}
\end{figure}

The quantification of the results were followed from the calculation of the Poincar\'{e} indices.
The values of $(SD1)$ and $(SD2)$ were calculated for lag$ = m$ from the relations
$SD1(m) = (\Phi(m) - \Phi(0))^{1/2}$ and $SD2(m) = (\Phi(m) + \Phi(0))^{1/2}$, where the auto-covariance function $\Phi(m)$ is given by
$\Phi(m) = E[(RR_{i} - RR_{M}) (RR_{i+m} - RR_{M})]$
and $RR_{M}$ is the mean of $RR_{i}$ \cite{Brennan},\cite{Bhaskar}.
We first analyze the results of indices for $m=1$ for three groups.

\begin{figure}[h!]
\centering
\includegraphics[width=0.8\textwidth]{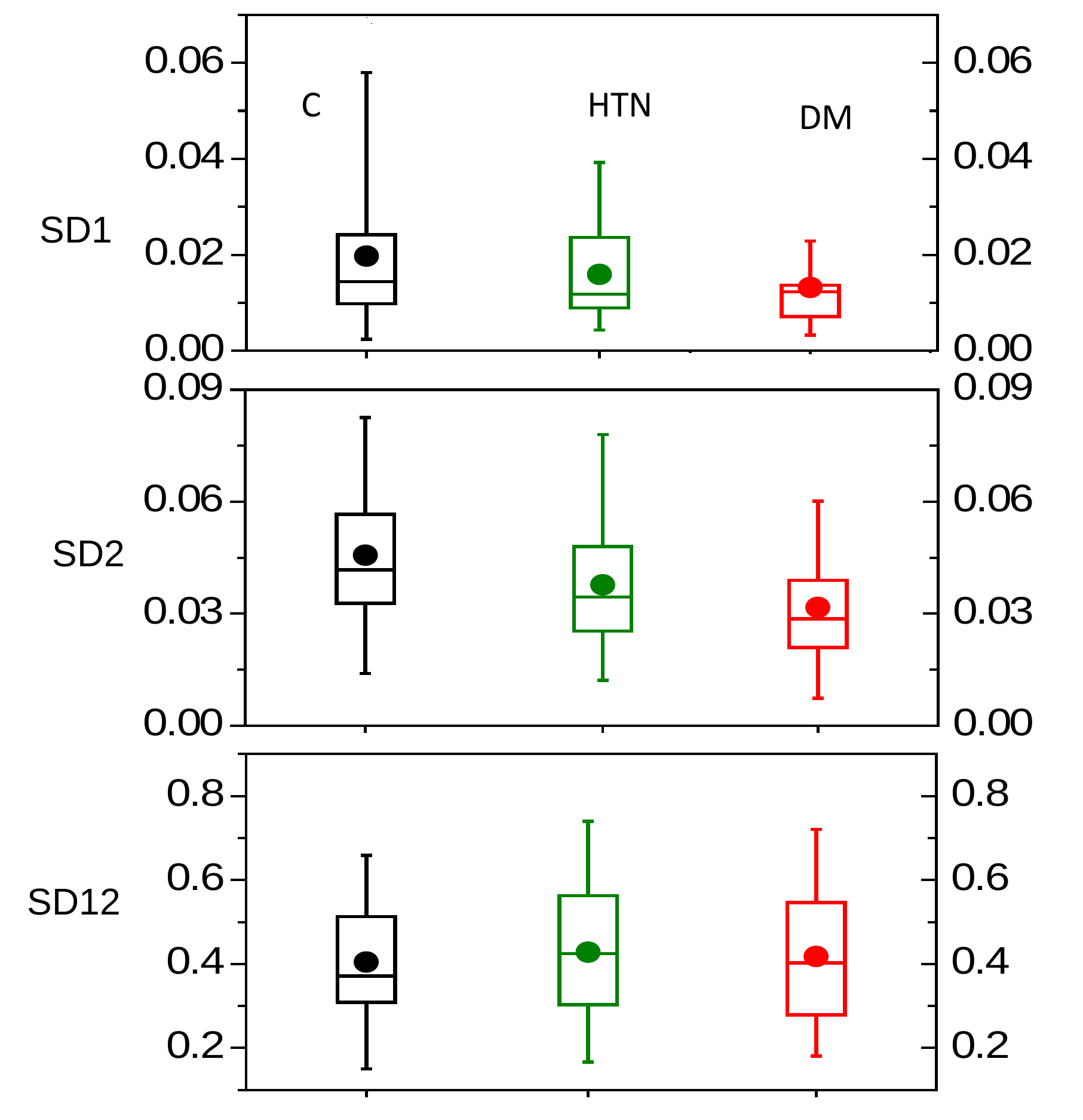}
\caption{The box plot of $(SD1)$,$(SD2)$ and $SD12$ for control (C),hypertensive (HTN) and diabetic (DM) subjects for lag$m=1$.The colour black,green and red are used through all figures for C,HTN and DM groups respectively.}
\label{sdbox}
\end{figure}

The  values $SD1$,$SD2$ and $SD12$ for each group for $m=1$ are presented in Fig.$2$ with mean (dot) ,median (line within box) and height of box marks $25,75 $ percent which is measure of distribution of data.The mean value of $SD1$ is lowest for DM group and that of HTN group lies in between value of groups C and DM.Similar results follow for $(SD2)$ albeit smaller differences. Moreover, the distribution of variability is also reduced with diabetic condition.The health conditions like diabetic and/or hypertension result shortening of both the short term and long term viabilities.As the short term variability is more affected,it is a better measure to monitor the HRV.On the other hand,the data on $SD12$ shows little contrast between the groups.However,the analysis of an extended Poincare\^{e} plot points out importance of $SD12$ and opens up for new measure of HRV.The mean values of $SD1$,$SD2$ and $SD12$ for different values of $m$ are obtained from $RR$ interval of individual subject for each group and their variation with $m$ are shown in Fig.$3$.

\begin{figure}[h!]
\centering
\includegraphics[width=0.5\textwidth]{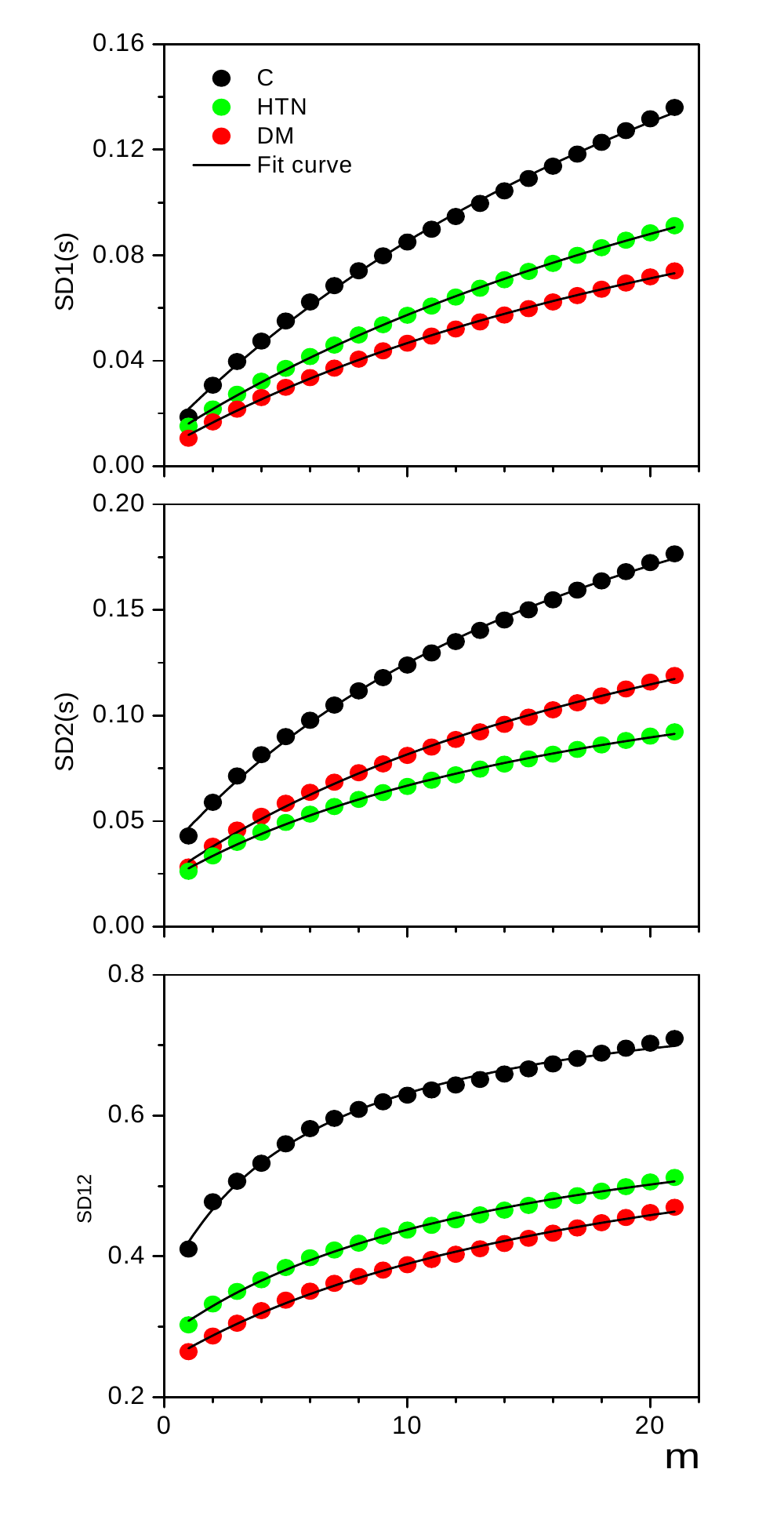}
\caption{The plots are the variation with $m$ of the mean values of $SD$ parameters for respective group.Points are measured values and the solid line is an approximate fitting.}
\label{allsd}
\end{figure}

The values of $SD1$ and $SD12$ are least for DM-group for all $m$.The subjects with HTN have least values for $SD2$.As expected,all three parameters resulting from the Poincar\'{e} plot for control group lies ahead of that of other group.All parameters for each group are an increasing function of lag variable $m$. We note that the growth rate of these parameters are higher close to the origin. This non-linear variation can be characterized by the slope and the curvature of plot.It is evident from scattered plot both the slope and the curvature decrease with $m$.For large value of $m$ the curvature becomes negligible as $SD12$ approaches to unity.To estimate the slope and the curvature we used the Pad\'{e} approximation \cite{Bhaskar}.A simple form for the Pad\'{e} approximation is chosen for analysis of non-linear variation of these parameters
\begin{equation}
\centering
 Y = \frac{a+bm}{1+cm}
 \label{pade}
\end{equation}
where $Y$ is either $SD1$,$SD2$ or $SD12$ and is represented by the ratio of linear polynomial of $m$ with three adjustable parameters $a$,$b$ and $c$.The data (scattered points)are fitted by varying the parameters and the solid curve (Fig.$3$) is best fitted curve with $R^2=99.9$ percent,and $ \chi^2 \sim 10^{-6} $.The slope and curvature of fitted curve had their large magnitude near $m=1$.
\begin{figure}[h!]
\centering
\includegraphics[width=0.8\textwidth]{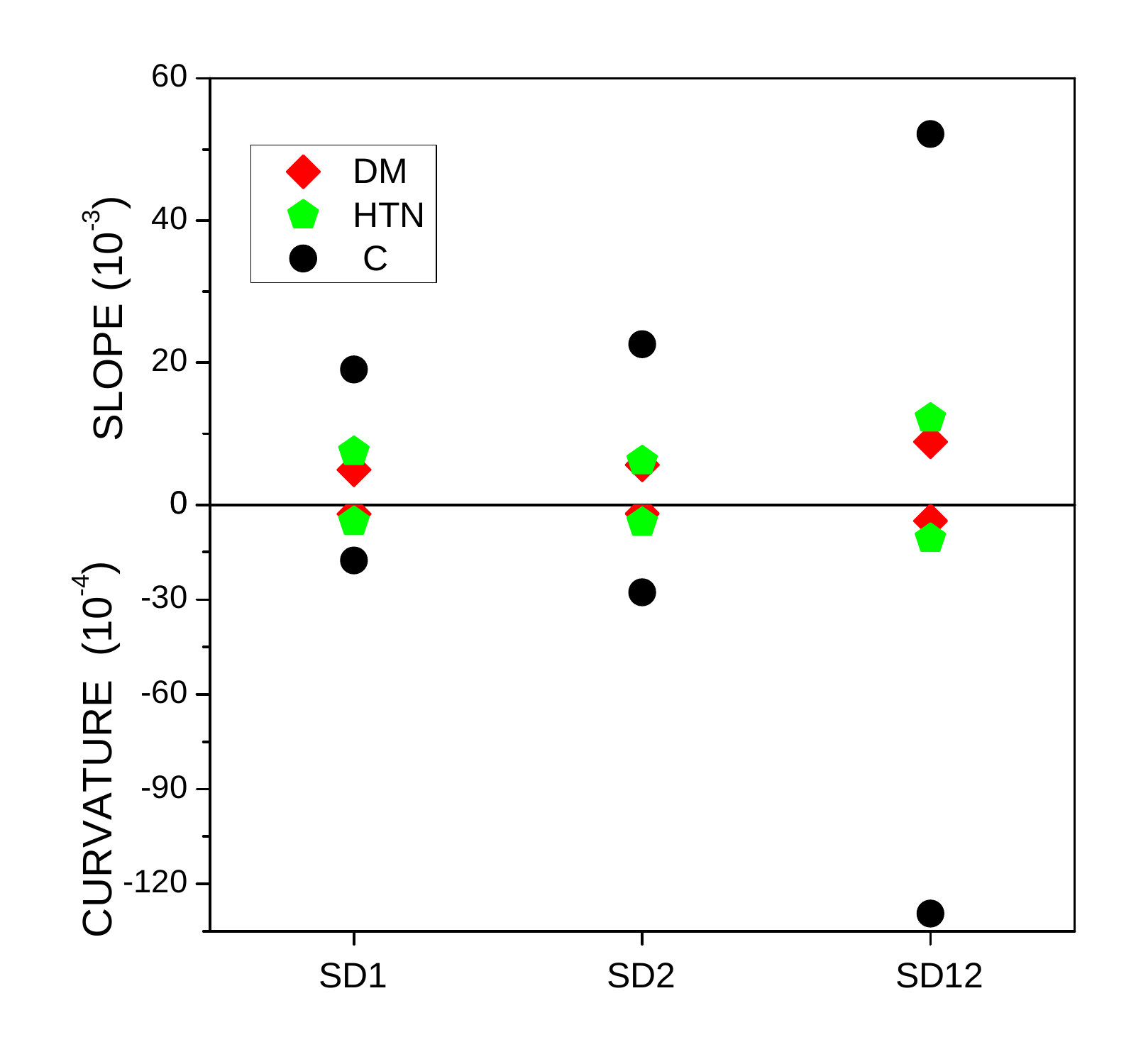}
\caption{The maximum value the slope and the curvature of plots obtained from fitted curve for each parameter and three groups of subjects.Note the scale difference}
\label{slopcurv}
\end{figure}
The maximum values of the slope and the curvature of each group as obtained from fitted equation are depicted in Fig.\ref{slopcurv} for all three parameters.Both the slope and the curvature of $SD12$ of control group are markedly different compared those other groups.In particular the value of the curvature for subjects with DM is much smaller.Note that the box plot of $SD12$ did not exhibit any distinguishable feature,the detail analysis of its variation with lag points out the important role in assessment of HRV.We next analyze in similar way the non-linear variation of the Poincar\'{e} parameters of all subjects using the Eq.\ref{pade}.The fitting of the data of three parameters of each subject is obtained with $R^2$ lying between $0.98 -0.99$.The slopes and the curvature are obtained from the fitted curve.It is observed that the slope and the curvature of $SD12$  exhibit larger variation compared to the results of other two parameters.The curvature is negative and the slope is of opposite sign for all cases.

\begin{figure}[h!]
\centering
\includegraphics[width=0.9\textwidth]{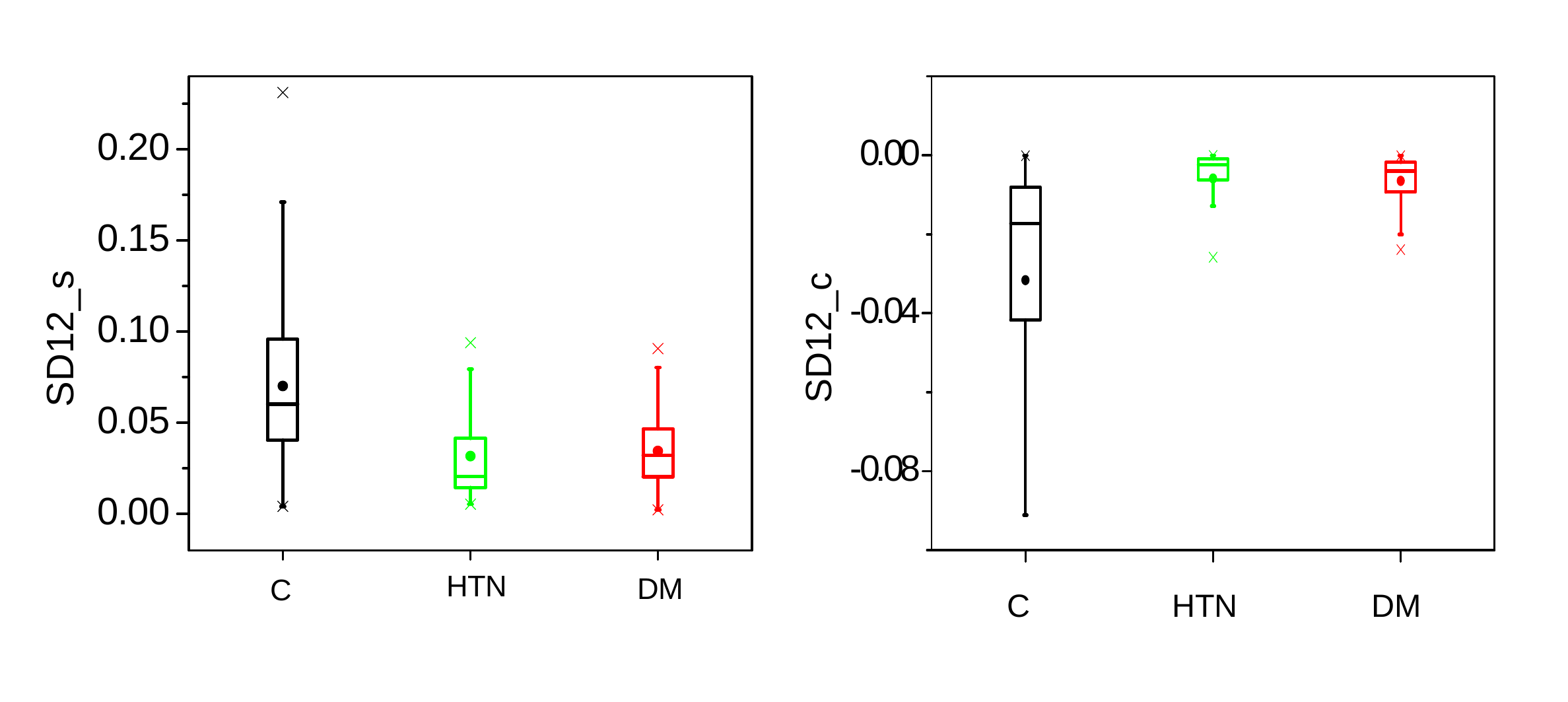}
\caption{The box plot of the slope $SD12$\textunderscore s (left) and the curvature $SD12$\textunderscore c(right) of $SD12$ of subjects of each group.}
\label{sl-cvbox}
\end{figure}
\begin{figure}[h!]
\centering
\includegraphics[width=1.0\textwidth]{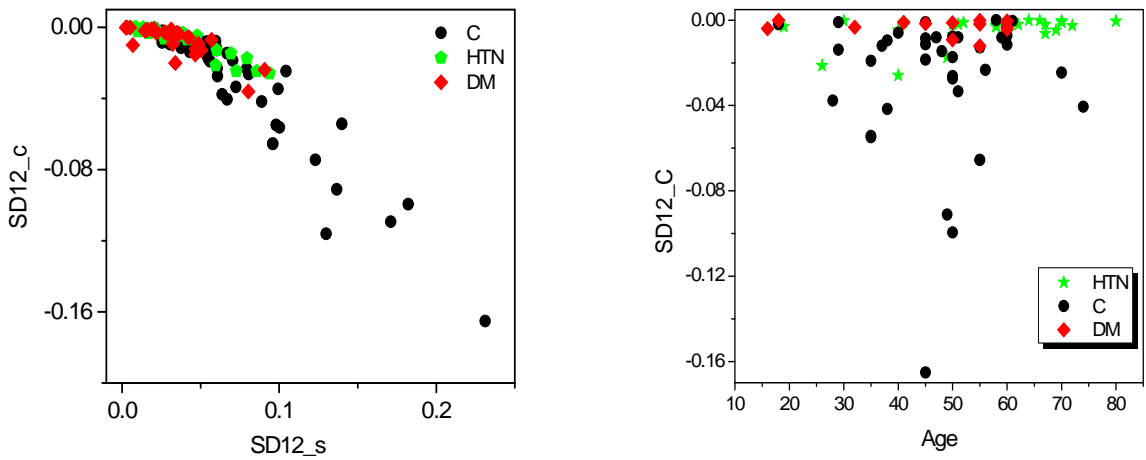}
\caption{ Plot of the curvature ($SD12$\textunderscore c) as a function of the slope ($SD12$\textunderscore s)(left) and as a function of age of the subjects(right) of $SD12$ (data of Fig.\ref{sl-cvbox}) for all three group of subjects.colour legend is same as in Fig.\ref{sl-cvbox}}
\label{curvage}
\end{figure}

The maximum values of the slopes $SD12$\textunderscore s and the curvature $SD12$\textunderscore c of $SD12 -m$ plot for individual subject within each group are considered.The results for control(C-black),for hypertensive (HTN-green)and diabetic (DM-red) are summarized in box-plot(Fig.\ref{sl-cvbox}).The magnitude of both theses quantities for control group are far greater than those of other two groups.This is a significant result.It was observed that the smaller value of the curvature was associated with the deviation of normal cardiovascular function of patient \cite{Thakre}.It is expected that the curvature tends to zero for heart without innervation and is large for heart of young person whose heart beat fluctuations are less correlated.We found that the young control subject had large value of curvature as shown in Fig.\ref{sl-cvbox}.Again,the spread of the variation of both the quantities are much smaller for the HTN- and DM- group.\\
Considering that the variation of $SD12$ with lag is more sharper, the values of the curvature $SD12$\textunderscore c of individual are replotted  against the respective value of slope $SD12$\textunderscore s (left figure)and the age (right figure)(Fig.\ref{curvage}).
Considering that the variation of $SD12$ with lag is more sharper the values of the curvature $SD12$\textunderscore c of individual are replotted  against the respective value of slope $SD12$\textunderscore s (left)and the age (right)(Fig.\ref{curvage}).
It appears that a power law relationship ($SD12$ \textunderscore c $\sim$ $SD12$ \textunderscore s$^{\alpha}$) exits however large scattering of data for control group makes $\alpha$ variable.Set with larger number of subject would help to establish the relationship. As regards the variation with age the control group has larger scattering compared with other two groups.For hypertensive subjects of older age group the curvature is small with very little variation.Similar results is found for diabetes.

\subsubsection*{\textit{ Principal Component Analysis of Poincar\'{e} Parameters}}

The parameters that come out from above analysis of the Poincar\'{e} plot are three $SD1$,$SD2$,$SD12$
and two quantities that defines their growth with $m$ namely-the maximum values of the slopes and the curvature.We have used the principal component analysis to envisage the relative importance of these $9$ parameters. For convenient presentation, the result the parameters $SD1$,maximum value of the slope $SD1$\textunderscore s and that of curvature $SD1$\textunderscore c are denoted by number $1$,$2$ and $3$ respectively, and similarly $4$ -$9$ refer to others in same sequence.The covariant matrix has been constructed with the normalized values following Eq.(\ref{matrix}) and the matrix element represents the correlation coefficients.
\begin{figure}[h!]
\centering
\includegraphics[width=1.0\textwidth]{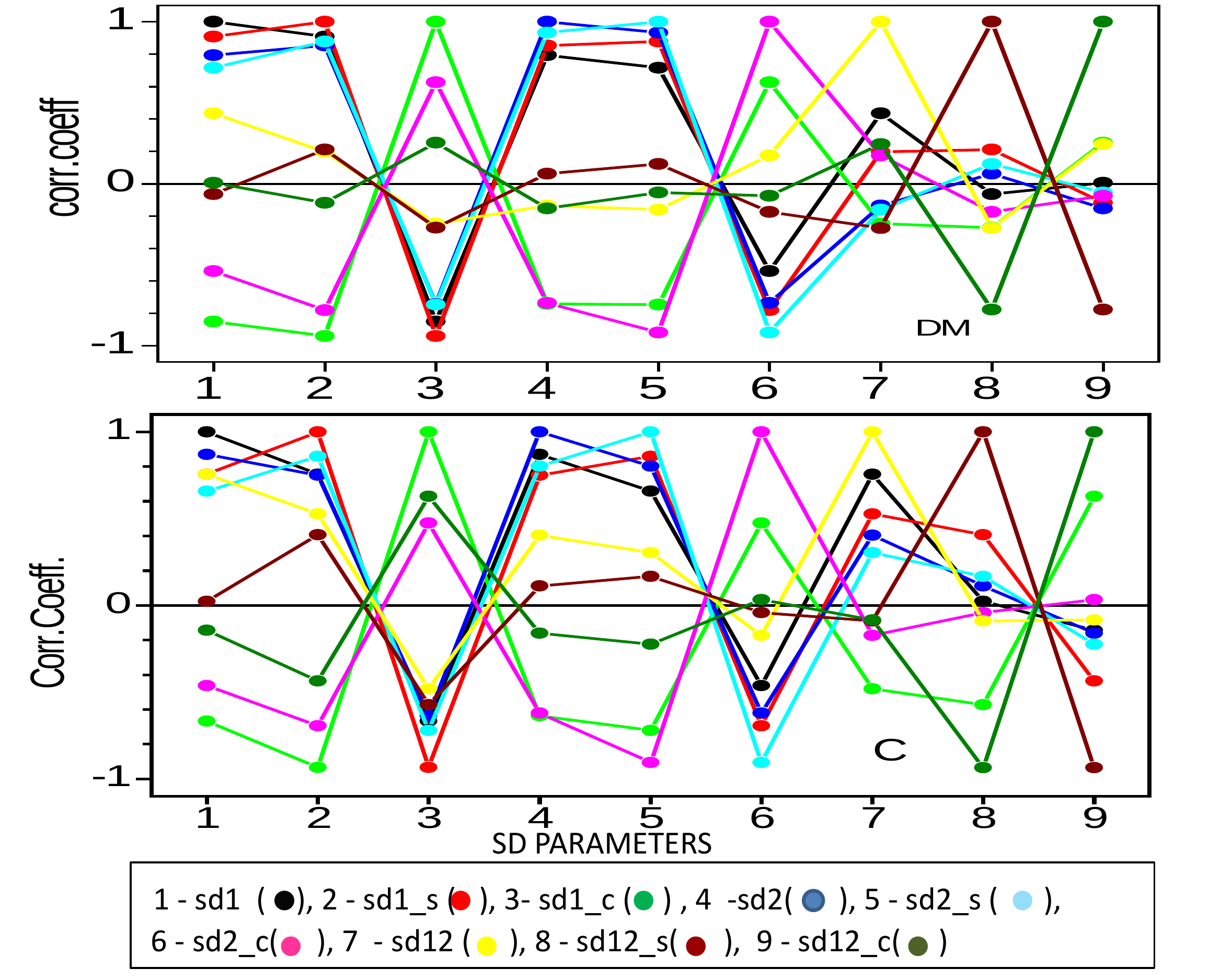}
\caption{ Plot of the correlation amplitudes which are  elements of covariant matrix of two groups control (C) and diabetic (DM).The X-axis number represents the SD-parameters, marked ($1-9$) as indicated below the figure.}
\label{corr}
\end{figure}
The auto-correlation for normalized parameters for the diabetic and control and  group is depicted in Fig.\ref{corr}.The slope and the curvature of a particular Poincar\'{e} parameter is highly correlated and negative value of the correlation coefficient is due to opposite sign of the slope and the curvature.The parameters $SD1$,$SD2$ are strongly correlated.On the other hand,$SD12$ is less dependent on other parameters.The slope and the curvature of $SD12$  are strongly correlated.The behaviour of the correlation coefficient for other group of subjects are similar in nature with smaller values of correlation coefficients.\\
The eigenvalues and eigenvectors \ref{eigvec} are found from diagonalization of the matrix $C$ and the results are summarized in figures given below.
\begin{figure}[h!]
\centering
\includegraphics[width=0.8\textwidth]{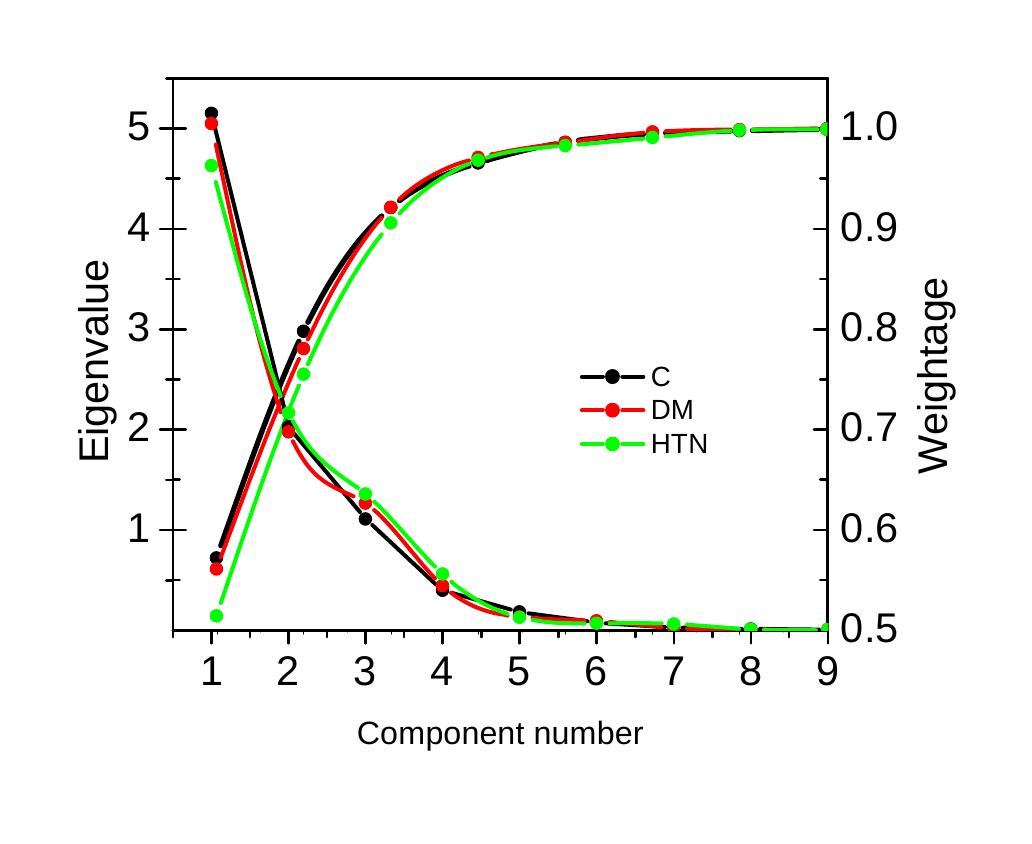}
\caption{ Plot of the eigenvalues ($\varepsilon_{i}$) and relative weight ($\sum _{1}^{i}\varepsilon_{i} /\sum_{i=1}^{9}\varepsilon_{i}$)  as a function of the SD-parameters.Legend of marking ($1-9$)is same as indicated earlier.}
\label{eigenvalu}
\end{figure}
The eigenvalues $\varepsilon_{i}$  of the correlation matrix exhibit identical trend for all three groups with highest value for control group Fig.\ref{eigenvalu}. Only few eigenvalues are of finite magnitude.Significance of eigenvalue is determined by its efficiency in retrieving data.The efficiency is usually estimated by examining by the relative weightage, defined as $\sum _{1}^{i}\varepsilon_{i} /\sum_{i=1}^{9}\varepsilon_{i}$.In present study the first four eigenvalues yield more than $90$ percent of the data as shown in Fig.\ref{eigenvalu}.This amounts to large reduction of dimension of data.
\begin{figure}[h!]
\centering
\includegraphics[width=0.9\textwidth]{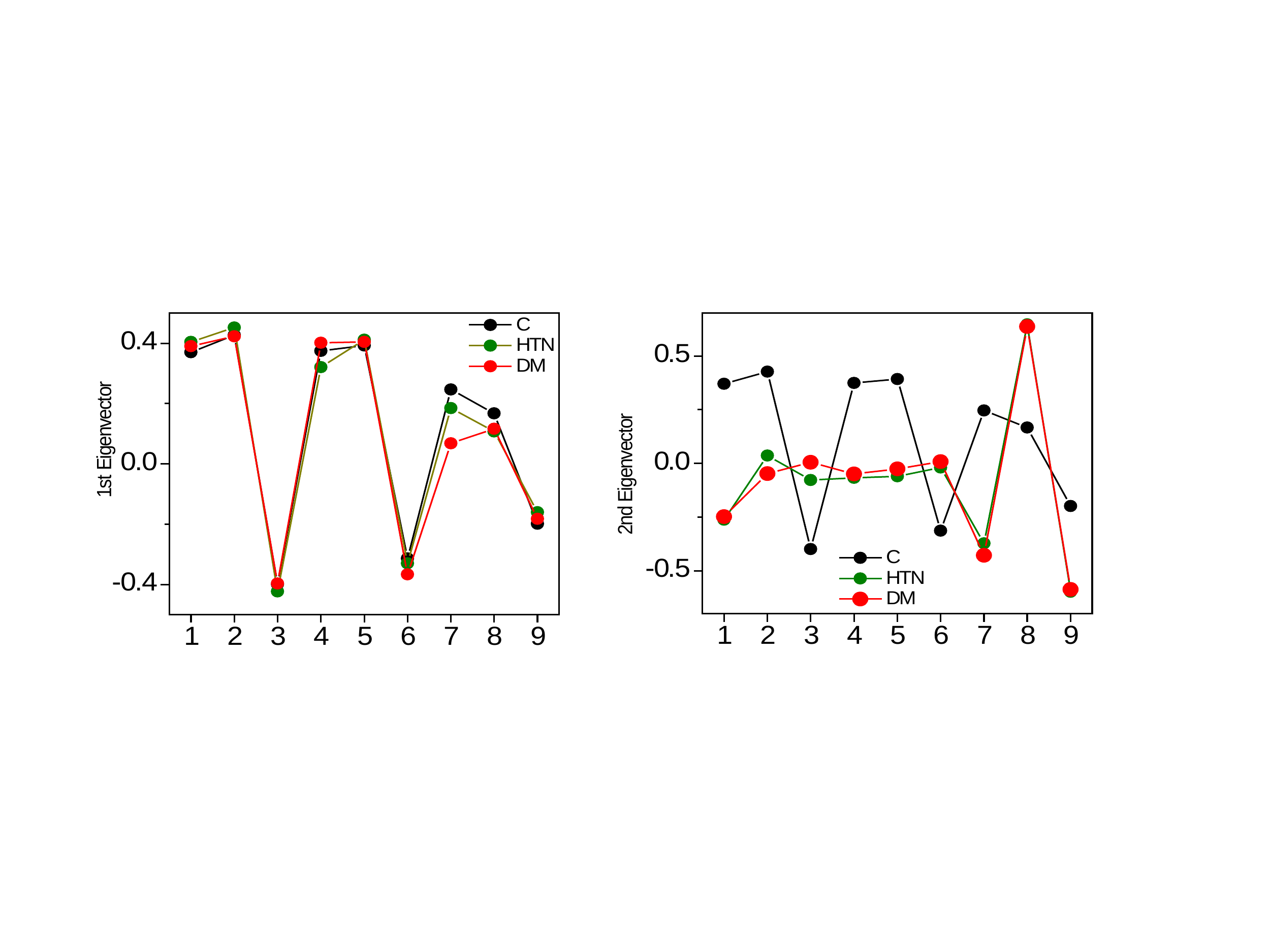}
\caption{ Plot of the first two eigenvectors corresponding to first two eigenvalues.Legend of marking ($1-9$) is same as noted earlier.}
\label{eigvec}
\end{figure}
The two significant eigenvectors (corresponding to two largest eigenvalues) for all three groups are shown in Fig.\ref{eigvec}.Nearly $75 - 80$ percent data are accounted by these vectors.Inclusion of third PCA gives more than $90$ percent data.The differences in value of first eigenvector for three groups are prominent for variable $SD12$.On the other hand values of second eigenvector are nearly alike for deceased subjects and differ significantly from the control group in all nine variables.The nature of variation of this vector for control group is similar to first one.It is evident that the curvature of plot $SD$'s -$m$ plays significant role in determining eigenvector.This in turn influences principal components.

\begin{figure}[h!]
\centering
\includegraphics[width=1.0\textwidth]{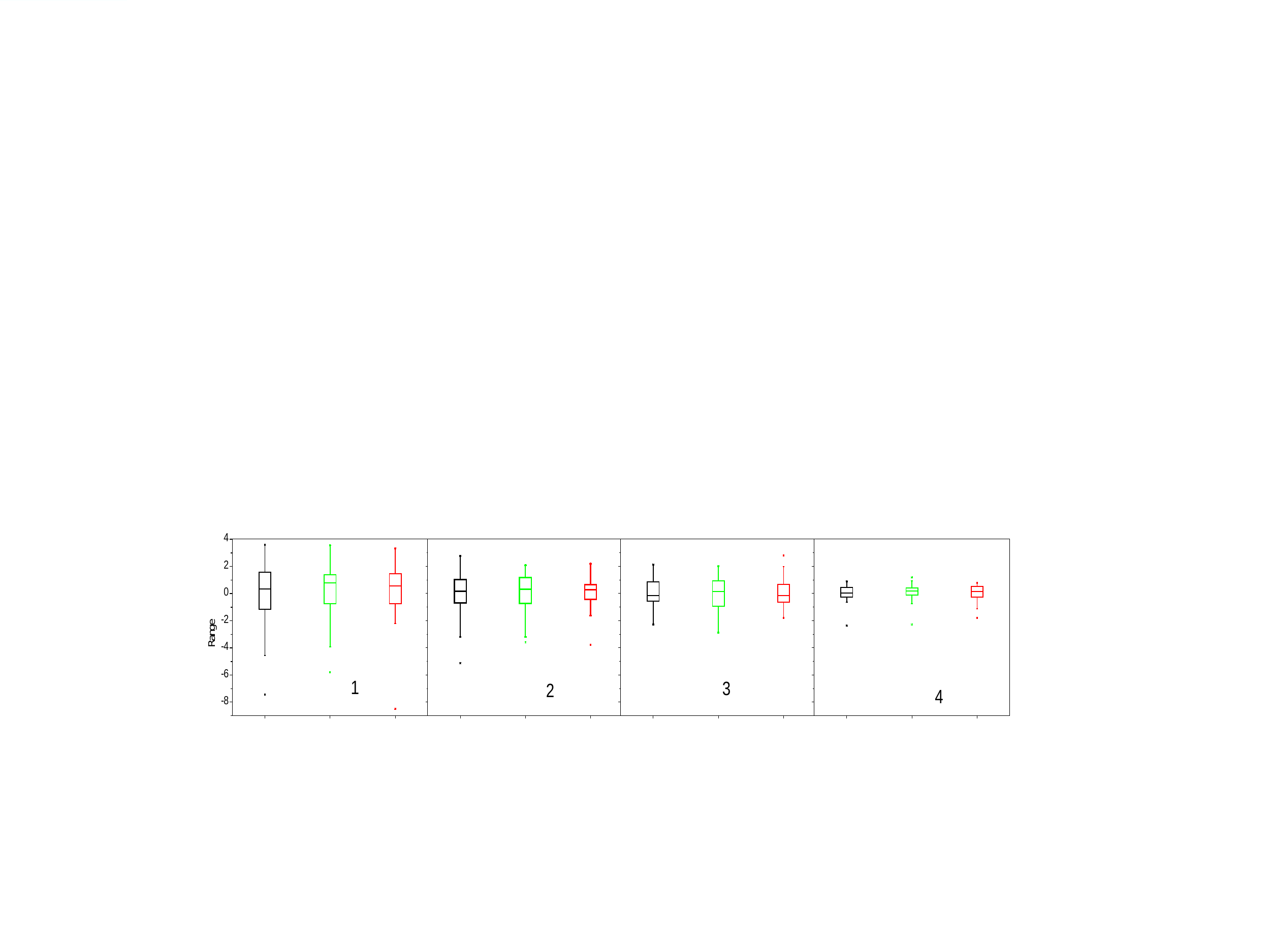}
\caption{Box plot of the first principal components(left figure),second and third ones (middle) and fourth one (right) for  the three groups (control C(black), hypertensive HTN(green) and diabetic DM(red)).The first principal component corresponds to highest eigenvalue and others are obtained from next three eigenvalues.}
\label{pcaplot}
\end{figure}
The first four principal components for three groups are depicted in the box-plot (Fig.\ref{pcaplot}).The difference among three groups is largest for first principal component.The width of distribution of the first principal component is large for control group and decreases successively for hypertensive and diabetic groups. Similarly,median values are wide apart.These differences in the variance of distribution of width for higher principal component number tend to vanish as shown in Fig.\ref{pcaplot}.






\begin{figure}[p]
    \makebox[\textwidth][c]{%
        \begin{minipage}[b]{1.0\linewidth}
            \centering
            \includegraphics[width=1.0\textwidth]{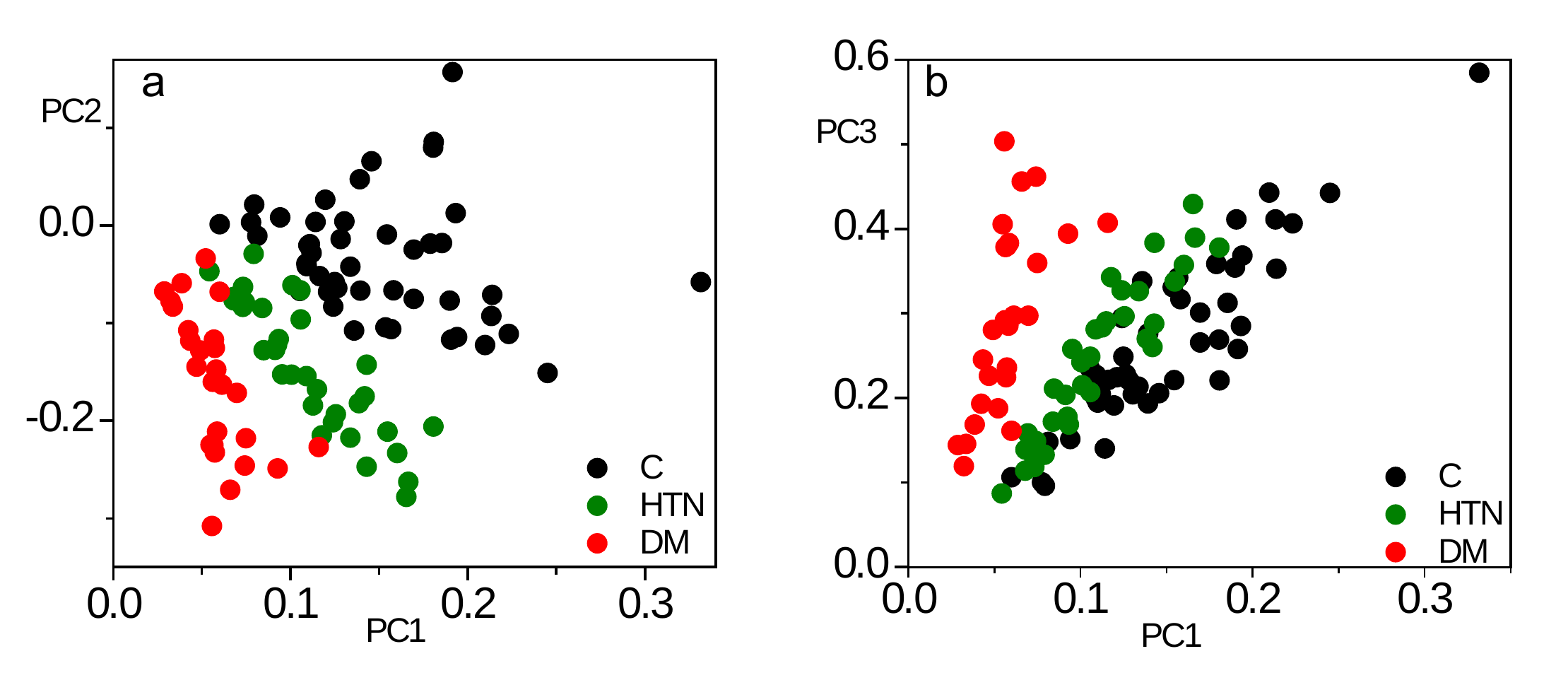}

        \end{minipage}}%

        \makebox[\textwidth][c]{%
        \begin{minipage}[b]{1.0\linewidth}
            \centering
            \hspace{-30pt}
            \includegraphics[width=1.02\textwidth]{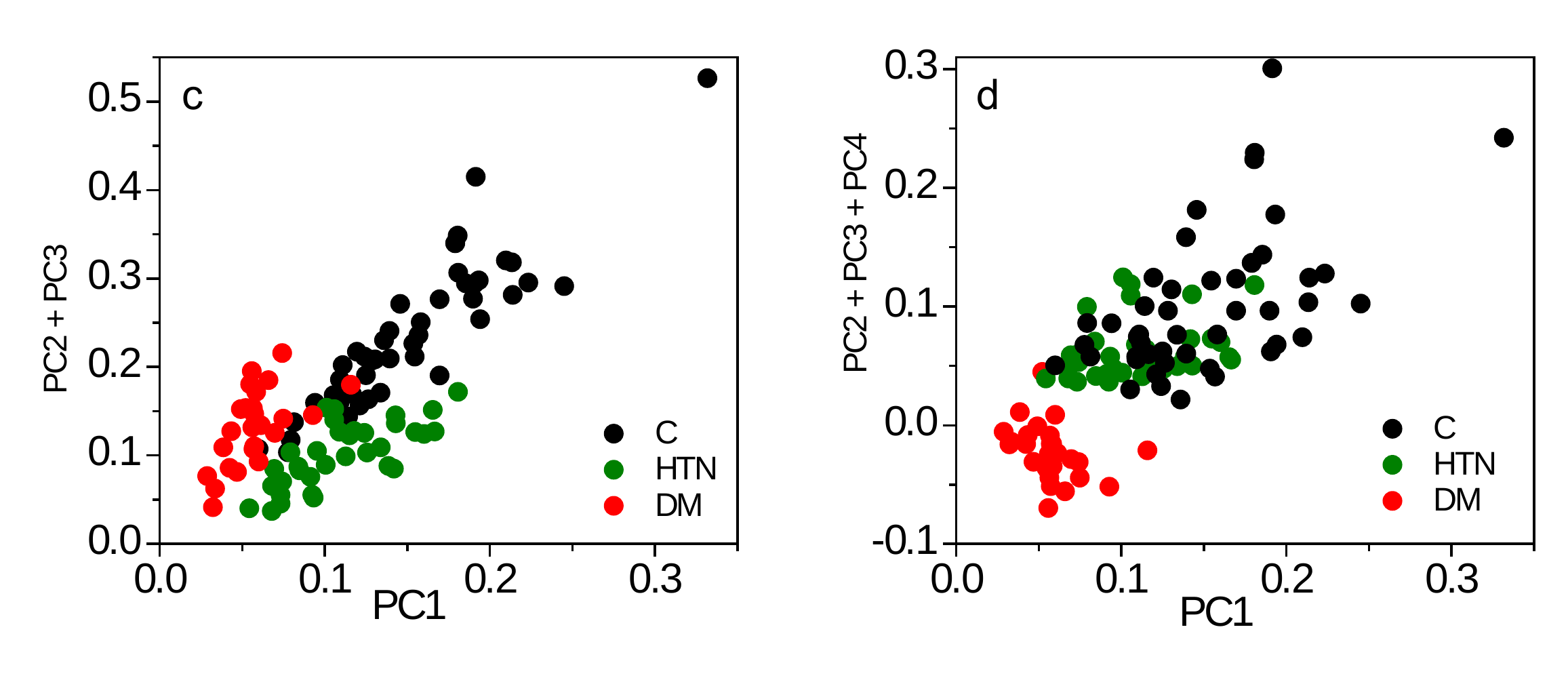}

        \end{minipage}}%
        \caption{Plot of first principal component $PC1$ vs \textit{a}) second component $PC2$,\textit{b}) third component $PC3$,\textit{c}) sum of $PC2$ and $PC3$ and \textit{d}) sum of $PC2$,$PC3$ and $PC4$ for subjects of each group(control C(black), hypertensive HTN(green) and diabetic DM(red))}
        \label{princomF}
        \end{figure}

The principal components corresponding to significant eigenvectors of individual subjects of each group are examined in Fig.\ref{princomF}.We analyse the result with respect to the first principal component $PC1$ corresponding to highest eigenvalue.The variation of other components (either single or combination) vis-vis first one are then expected to elucidate importance of PCA.The second $PC2$ and third $PC3$ component variation are given in Fig.\ref{princomF} (\textit{a}) and (\textit{b}) respectively.Similarly, the plot (\textit{c}) and (\textit{d}) of Fig.\ref{princomF} depict respectively the variation of sum ($PC2+PC3$) and ($PC2+PC3+PC4$) with $PC1$.First thing to note is the concentration of data into different region in phase space of principal components.The data for each group are well separated as is evident in Fig.\ref{princomF} (\textit{a}) and (\textit{c}).The data of hypertensive group lies in the middle of data for control and diabetic groups (Fig.\ref{princomF}\textit{a}).In (Fig.\ref{princomF}\textit{c}) the data of each group have distinct region of phase space with very little overlap.On the other hand,there is considerable overlap of points for control and hypertensive groups when the plot of third principal component (Fig.\ref{princomF}\textit{b}) or ($PC2+PC3+PC4$) (Fig.\ref{princomF}\textit{d}) against first one.However,the points for diabetic group is well-separated from those of other group.

\subsubsection*{\textit{Correlation of Fluctuation of Beat Interval}}
The correlation function ($Corr$) as defined in Eq.(\ref{cor}) is plotted as a function of lag $m$ for one subject from each group.The subjects are young with similar age.In order to examine extent of correlation the $RR$ intervals were randomised by shuffling successively $5$ times.Further shuffling did not change character of correlation of shuffled data.
\begin{figure}[!htb]
\minipage{0.32\textwidth}
  \includegraphics[width=1.1\linewidth]{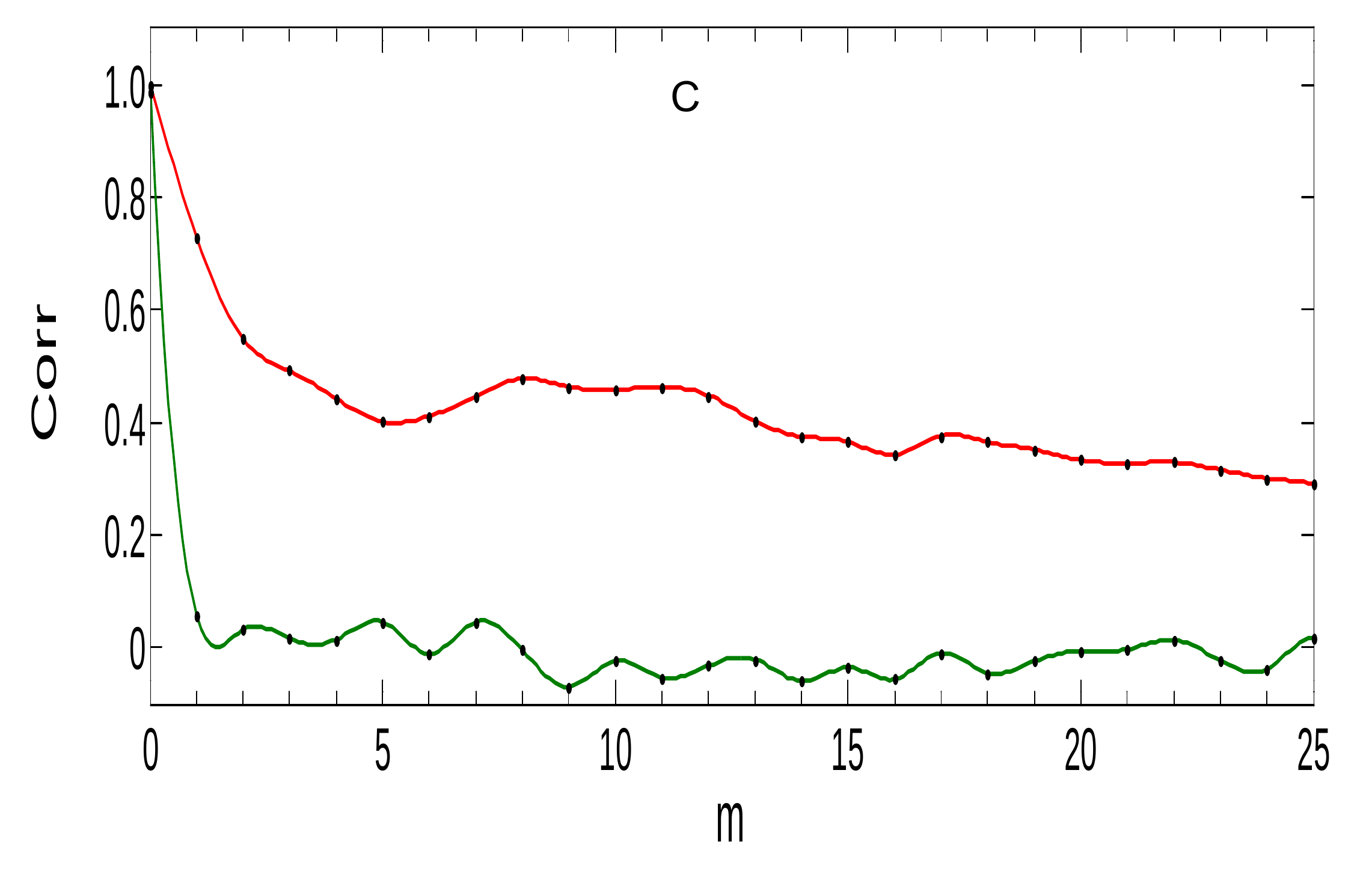}
\endminipage
\hfill
\minipage{0.32\textwidth}
 \includegraphics[width=1.1\linewidth]{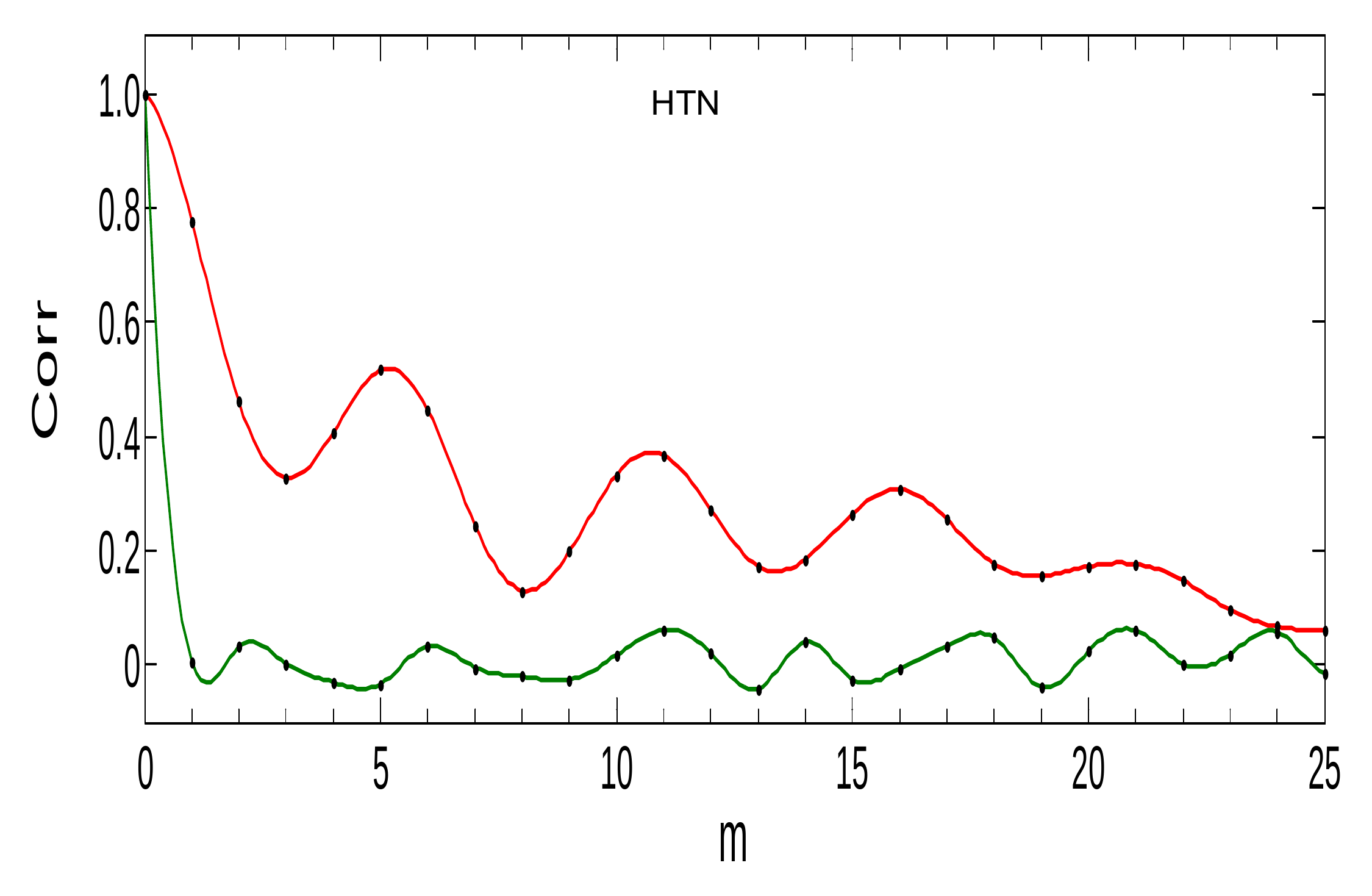}
\endminipage\hfill
\minipage{0.32\textwidth}%
 \includegraphics[width=1.1\linewidth]{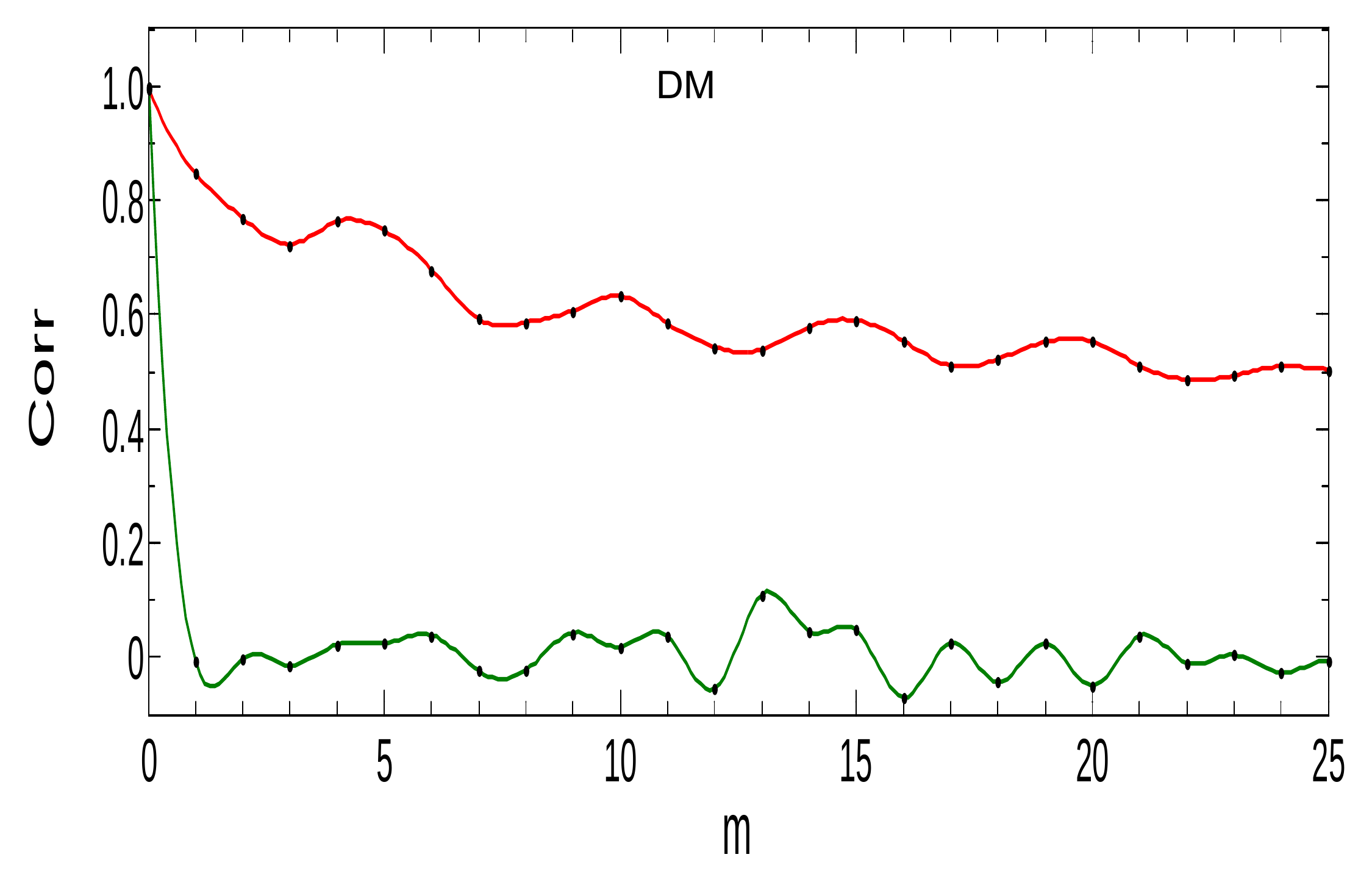}
\endminipage
\caption{Correlation function $Corr$ of $RR$ interval fluctuation as a function of lag $m$.The points (dot) are calculated data and the continuous curves are line through points.The red curve is for calculated data and green one for shuffle data}
\end{figure}
The figure marked by C ,HTN and DM are respectively for control,hypertensive and diabetic group.The red line are curve through calculated points and green one is that through shuffle points.The correlation of shuffled data are all alike and exhibit random character with near-zero correlation for finite $m$.The small fluctuation around zero is related to error. On the other hand,the correlation of $RR$ fluctuation decays with $m$ and remains finite for appreciable lag.The correlation function for subject of HTN/DM are oscillatory in nature.The decay of correlation for HTN case is faster than that for DM one.The oscillatory decay of the correlation for young control subject is not apparent.However,for older subject from same group such oscillation exists.

\subsection*{Discussions and Conclusions}
The differences in HRV pattern for three groups were exemplified from the detail and nonlinear analysis of the Poincar\'{e} parameters and their derivatives.The changes of cardiovascular regulation due to health disorder are reflected in the difference in distribution of parameters.The lagged Poincar\'{e} plot is found to reveal more effectively the differences in HRV \cite{Bhaskar}.It has been recognized that nearly eight consecutive of heart beat are influenced by any given RR interval and this notion triggers the analysis of lagged plot \cite{Thakre},\cite{Lerma}.As $SD1$ correlates with short term variability of heart rate and is mainly determined by parasympathetic response,the smaller
value of $SD1$ for diseased groups indicates weakening of parasympathetic regulation by health disorder\cite{Bhaskar}.The genesis of hypertension is a complex interaction of multiple pathophysiologic factors that include increased sympathetic nervous system acivity.It contributes to both the development and maintenance of hypertension through stimulation of the heart,peripheral vasculature and kidneys.Autonomic imbalance(increased sympathetic and decreased parasympathetic tone)is known to be associated with increased cardiovascular morbidity and mortality.The
$SD2$ (long term variability) is related more strongly to sympathetic activity than parasympathetic activity. When the $SD2$ value decreases,sympathetic activity is increased.The Poincar\'{e} parameters are increasing function of lagged value $m$.For large value of $m$, $SD2$ tends to unity as beats very well separated  in time are not correlated.Important characteristics of growth of these indices turn out to be slope and curvature for low value of $m$.Out of these six derived quantities the slope and -in particular curvature of $SD12$ of three groups differ significantly.The reduction of the $SD12$\textunderscore c with age of the subject is evident indicating imbalanced ANS activity.The principal component analysis of the variables provides more information embedded in the lagged Poincar\'{e} plots.The first four PC associated with four significant eigenvalues of the covariant matrix of data preserve most of the variability of data.The trajectory of PC when plotted with appropriate combination of PC clearly separates out three group.It also turns out that the PC which is weighted sum of original variables is dominated by the variable belonging to $SD12$.The correlation of beat interval fluctuation shows that the heart beats are far from random in character.Oscillatory character of the correlation function is associated with health disorder.The more detail analysis of correlation will be treated later in future.

In conclusion, the comparative strength of the Poincar\'{e} indices and their growth with lag index
and thereafter the principal component analysis might be useful to distinguish normal from pathological HRV. However,our results need to be validated by a larger representative number of subjects. Additionally, other studies, including pharmacological and physiological interventions that affect the autonomic nervous system could be used to test hypotheses that may establish our findings.

\subsection*{Acknowledgement}
The assistance and help from SSKM Hospital,Kolkata and Ghatmura Vivekananda Pathachkra are thankfully acknowledged.We also thankful to Dr.A.C. Roy  and Dr.Nirmalendu Sarkar for assistance and encouragement.SKG is grateful to Divyananda Maharaj in helping me to serve Vidyamandir,Belur
\subsection*{Declaration of Interest}
The authors declared no biomedical financial interest or potential conflicts of interest.

\end{document}